\documentclass[aps,prb,amsmath,amssymb,superscriptaddress,twocolumn,longbibliography]{revtex4-2}
\usepackage{bbm}
\usepackage{graphicx}
\usepackage{dcolumn}
\usepackage{bm}
\usepackage{subfigure}
\usepackage{amsmath}
\usepackage{stmaryrd}
\usepackage{times}
\usepackage{graphicx}
\usepackage{dcolumn}
\usepackage{color}
\usepackage{makecell}
\usepackage{tabularx,multirow}
\usepackage{appendix}
\usepackage{amssymb}
\usepackage{pifont}
\usepackage[colorlinks,linkcolor=blue,citecolor=blue,urlcolor=black]{hyperref}

\begin{document}
\title{Magnetic control of nonlinear transport induced by the quantum metric}
\author{Xu Chen}
\affiliation{School of Physics, Harbin Institute of Technology, Harbin 150001, China}
\author{Mingbo Dou}
\affiliation{School of Physics, Harbin Institute of Technology, Harbin 150001, China}
\author{Qin Zhang}
\affiliation{School of Physics, Harbin Institute of Technology, Harbin 150001, China}
\author{Xianjie Wang}
\affiliation{School of Physics, Harbin Institute of Technology, Harbin 150001, China}
\affiliation{Frontiers Science Center for Matter Behave in Space Environment, Harbin Institute of Technology, Harbin 150001, China}
\affiliation{Heilongjiang Provincial Key Laboratory of Advanced Quantum Functional Materials and Sensor Devices, Harbin 150001, China}
\author{M. Ye. Zhuravlev}
\affiliation{Faculty of Liberal Arts and Sciences, St. Petersburg State University, St. Petersburg 190000, Russia}
\author{A. V. Nikolaev}
\affiliation{Skobeltsyn Institute of Nuclear Physics, Moscow State University, Moscow 101000, Russia}
\author{L. L. Tao}
\email{Contact author: lltao@hit.edu.cn}
\affiliation{School of Physics, Harbin Institute of Technology, Harbin 150001, China}
\affiliation{Frontiers Science Center for Matter Behave in Space Environment, Harbin Institute of Technology, Harbin 150001, China}
\affiliation{Heilongjiang Provincial Key Laboratory of Advanced Quantum Functional Materials and Sensor Devices, Harbin 150001, China}
\date{\today}
\begin{abstract}
The quantum geometry plays a crucial role in the nonlinear transport of quantum materials. Here, we use the Boltzmann transport formalism to study the magnetic control of nonlinear transport induced by the quantum metric in two-dimensional systems with different types of spin-orbit coupling (SOC). It is shown that the nonlinear conductivity is strongly dependent on the direction of a field and reveals significant spatial anisotropy. Moreover, the field-direction dependent relations are distinct for different SOCs. In addition, it is demonstrated that the contributions from the quantum metric and Drude mechanism are distinguishable due to their opposite signs or distinct anisotropy relations. We further derive the analytical formulas for the anisotropic nonlinear conductivity, in exact agreement with numerical results. Our work shines more light on the interplay between the nonlinear transport and quantum geometry.
\end{abstract}
\maketitle
\section{Introduction}
The nonlinear transport due to the \emph{intrinsic} quantum geometry of wave functions has attracted growing interest\cite{Wilczek,rmd1959,nrp744,prl240001,nsrnwae334,nm2025}. For example, it was predicted\cite{prl216806,prb041101} that the Berry curvature dipole (BCD) in $\mathcal{T}$-invariant ($\mathcal{T}$ for time reversal) materials can induce a nonlinear anomalous Hall effect (NLAHE), which was experimentally observed in Weyl semimetals WTe$_2$ and MoTe$_2$\cite{nature337,nm324,nc2049}. In addition, the \emph{intrinsic} NLAHE can be induced by the quantum metric\cite{prl166601}, as demonstrated in antiferromagnets CuMnAs\cite{prl277201}, Mn$_2$Au\cite{prl277202}, MnBi$_2$Te$_4$\cite{sci181}, manganese chalcogenides MnX (X=S, Se, Te) monolayer\cite{prl056401}, and the altermagnet RuO$_2$\cite{prl106701}. Inversely, measuring the NLAHE is useful to probe the quantum geometry in quantum materials and can also be exploited to detect the direction of N\'eel vector in antiferromagnets\cite{prl277201,prl277202,prl056401}. Recently, the displacement current under an AC electric field driven by the quantum metric was demonstrated and its quantum theory was formulated\cite{prb115121}.

It is known that the second-order nonlinear conductivity $\sigma^{\text{(2)}}_{abc}$ ($a, b, c=x, y, z$ for Cartesian components) characterizing the nonlinear transport has three contributions\cite{prl026301,prbL201405,prbL241405}: the nonlinear Drude conductivity $\sigma^{\text{Drude}}_{abc}$ caused by the band asymmetry, the BCD induced nonlinear conductivity $\sigma^{\text{BCD}}_{abc}$ and the quantum metric induced nonlinear conductivity $\sigma^{\text{QM}}_{abc}$. Importantly, $\sigma^{\text{BCD}}_{abc}$ represents a purely transverse effect while $\sigma^{\text{Drude}}_{abc}$ and $\sigma^{\text{QM}}_{abc}$ have both transverse and longitudinal effects due to their identical symmetry restrictions\cite{prl026301,prbL201405,prbL241405}. In earlier studies\cite{prl166601,prl277201,prl277202}, it was indicated that $\sigma^{\text{QM}}_{abc}$ contributes only to the transverse response while the longitudinal component is vanishing. Recent works\cite{prl026301,prbL201405} show that $\sigma^{\text{QM}}_{abc}$ also contributes the longitudinal nonlinear transport and the quantum metric dominated nonreciprocal charge transport was observed in MnBi$_2$Te$_4$ thin films\cite{na487}. On the other hand, $\sigma^{\text{Drude}}_{abc}$, $\sigma^{\text{BCD}}_{abc}$ and $\sigma^{\text{QM}}_{abc}$ have distinct relaxation time $\tau$ dependent relations: $\sigma^{\text{Drude}}_{abc}$ and $\sigma^{\text{BCD}}_{abc}$ are $\tau^2$ and $\tau$ dependent, respectively, while $\sigma^{\text{QM}}_{abc}$ is $\tau$ independent. Thus, $\sigma^{\text{QM}}_{abc}$ dominates the nonlinear transport in the dirty limit ($\tau\rightarrow0$) and represents an \emph{intrinsic} contribution to the nonlinear transport.

The nonreciprocal charge transport (NCT)\cite{sae1602390,nc3740,nrp558} is characterized by different resistances for opposite currents and can be described by the longitudinal nonlinear conductivity $\sigma^{\text{(2)}}_{aaa}=\sigma^{\text{Drude}}_{aaa}+\sigma^{\text{QM}}_{aaa}$ due to the vanishing of $\sigma^{\text{BCD}}_{aaa}$. Both $\mathcal{P}$ ($\mathcal{P}$ for inversion) and $\mathcal{T}$ symmetries must be broken to obtain nonzero $\sigma^{\text{Drude}}_{abc}$ or $ \sigma^{\text{QM}}_{abc}$. Indeed, the sizable NCT effect due to $\sigma^{\text{Drude}}_{abc}$ was observed in ferromagnetic/antiferromagnetic materials\cite{prb054429,prl276601,prl096802} and non-magnetic materials without $\mathcal{P}$ symmetry by applying an external magnetic field\cite{np578,nc540,prr033253,prb115202,prl176602,prl046303,prr013041}. In those systems, the band asymmetry responsible for $\sigma^{\text{Drude}}_{abc}$ is caused by the combined spin-orbit coupling (SOC) and magnetic field. Recently, we studied the anisotropic nonlinear Drude conductivity based on the Boltzmann transport theory\cite{prb155411}. It was shown that the nonlinear Drude conductivity reveals significant spatial anisotropy and different SOCs reveal distinct field-direction dependent relations. In this work, we focus on the anisotropic nonlinear conductivity induced by the quantum metric.

The rest of the paper is organized as follows. In Sec. \ref{sec2}, we present the theoretical formalism and general formula for the nonlinear conductivity calculations. In Sec. \ref{sec3}, we discuss the magnetically tunable anisotropic nonlinear conductivity based on the general Hamiltonian model. Finally, Sec. \ref{sec4} is reserved for further discussion and conclusion.

\section{Theoretical formalism\label{sec2}}

\begin{table*}
\caption{\label{table1} The derived $\sigma_{\phi\phi\phi}^{\text{Drude}}$ and $\sigma_{\phi\phi\phi}^{\text{QM}}$ for different SOCs within the weak-field or high-density regime. $\mathbf{\Omega(k)}$ is the corresponding spin-orbit field and $\alpha, \beta, \gamma, \delta$ are the SOC parameters. The prefactors $\sigma^{\text{Drude}}_{R/D/W/P}$ are given in Ref. \cite{prb155411} while $\sigma^{\text{QM}}_{R/D/W/P}$ are given in the main text.}
\begin{ruledtabular}
\begin{tabular}{ccccc}
SOC & $\mathbf{\Omega(k)}$ & $\sigma_{\phi\phi\phi}^{\text{Drude}}\cite{prb155411}$ & $\sigma_{\phi\phi\phi}^{\text{QM}}$\\
\hline
RSO & $\alpha(-k_y, k_x, 0)$ & $\sigma^{\text{Drude}}_R\sin\theta\sin(\varphi-\phi)$ & $\sigma^{\text{QM}}_R\sin\theta\sin(\varphi-\phi)$ \\
DSO & $\beta(k_x, -k_y, 0)$ & $\sigma^{\text{Drude}}_D\sin\theta\cos(\varphi+\phi)$ & $\sigma^{\text{QM}}_D\sin\theta\cos(\varphi+\phi)$ \\
WSO & $\gamma(k_x, k_y, 0)$ & $\sigma^{\text{Drude}}_W\sin\theta\cos(\varphi-\phi)$ & $\sigma^{\text{QM}}_W\sin\theta\cos(\varphi-\phi)$ \\
PSO & $\delta(k_x-k_y, k_x-k_y, 0)$ & $\sigma^{\text{Drude}}_P\sin\theta\sin(\varphi+\frac{\pi}{4})\sin(\phi-\frac{\pi}{4})$ & $\sigma^{\text{QM}}_P\sin\theta\sin(\varphi+\frac{\pi}{4})\sin^3(\phi-\frac{\pi}{4})$ \\
\end{tabular}
\end{ruledtabular}
\end{table*}

The starting point is the following Hamiltonian describing the two-dimensional (2D) system with SOC and Zeeman effects\cite{prb155411,npj172}:
\begin{equation}\label{eq-1}
    \mathcal{H}=\frac{\hbar^2k^2}{2m_e}+\mathbf{\Omega(k)}\cdot\mathbf{\sigma}-\Delta\hat{\mathbf{m}}\cdot\mathbf{\sigma},
\end{equation}
The first term represents the kinetic energy given in terms of electron effective mass (isotropic approximation) $m_e$, the reduced Planck's constant $\hbar$, and the wave vector $\mathbf{k}=(k_x, k_y)=k(\text{cos}\phi, \text{sin}\phi)$ in the Cartesian and polar coordinates ($\phi$ for azimuthal angle). The second term describes the SOC with $\mathbf{\Omega(k)}=(\Omega_x, \Omega_y, \Omega_z)$ and $\mathbf{\sigma}=(\sigma_x, \sigma_y, \sigma_z)$ being the spin-orbit field and the vector of Pauli matrices, respectively. The third term is the Zeeman term, where $\Delta$ characterizes the magnitude of the exchange field or an external magnetic field and the unit vector $\hat{\mathbf{m}}=(\text{sin}\theta\text{cos}\varphi, \text{sin}\theta\text{sin}\varphi, \text{cos}\theta)$ ($\theta$ for polar angle and $\varphi$ for azimuthal angle) denotes the direction of a field. It is noteworthy that the exchange field can be induced by a ferromagnetic or antiferromagnetic insulator substrate through the magnetic proximity effect\cite{mt85}. Equation (\ref{eq-1}) can be rewritten as $\mathcal{H}=\hbar^2k^2/(2m_e)+\mathbf{h(k)}\cdot\mathbf{\sigma}$ with $\mathbf{h(k)}\equiv\mathbf{\Omega(k)}-\Delta\hat{\mathbf{m}}$ being the effective field. The eigenvalues $\epsilon_{\mathbf{k}s}$ ($s=\pm1$ for spin index) and normalized eigenstates $\psi_{\mathbf{k}s}$ can be obtained as\cite{prb085438}
\begin{equation}\label{eq-2}
\begin{aligned}
  &\epsilon_{\mathbf{k}s}=\frac{\hbar^2k^2}{2m_e}+sh,\\
  &\psi_{\mathbf{k}s}=\frac{e^{i\mathbf{k}\cdot\mathbf{r}}}{\sqrt{2h(h-sh_z)}}\left(
                                                                                   \begin{array}{c}
                                                                                     h_x-ih_y \\
                                                                                     sh-h_z \\
                                                                                   \end{array}
                                                                                 \right).
\end{aligned}
\end{equation}
In this work, we consider four different SOCs, that is, Rashba (RSO)\cite{rashba1109,am509,prb220101}, Dresselhaus (DSO)\cite{prb580,prb245159,prb245141}, Weyl (WSO)\cite{prl136404,prl216402,afm2208023}, and the SOC with persistent spin texture (PSO)\cite{prl236601,rmp011001,pst073002,nc2763,apl022411,2d025026,prl076801} as similar to previous works\cite{prb085438,jpd113001,njp123005} and the corresponding $\mathbf{\Omega(k)}$ is listed in Table \ref{table1}.

\begin{figure*}
\includegraphics[width=0.9\textwidth]{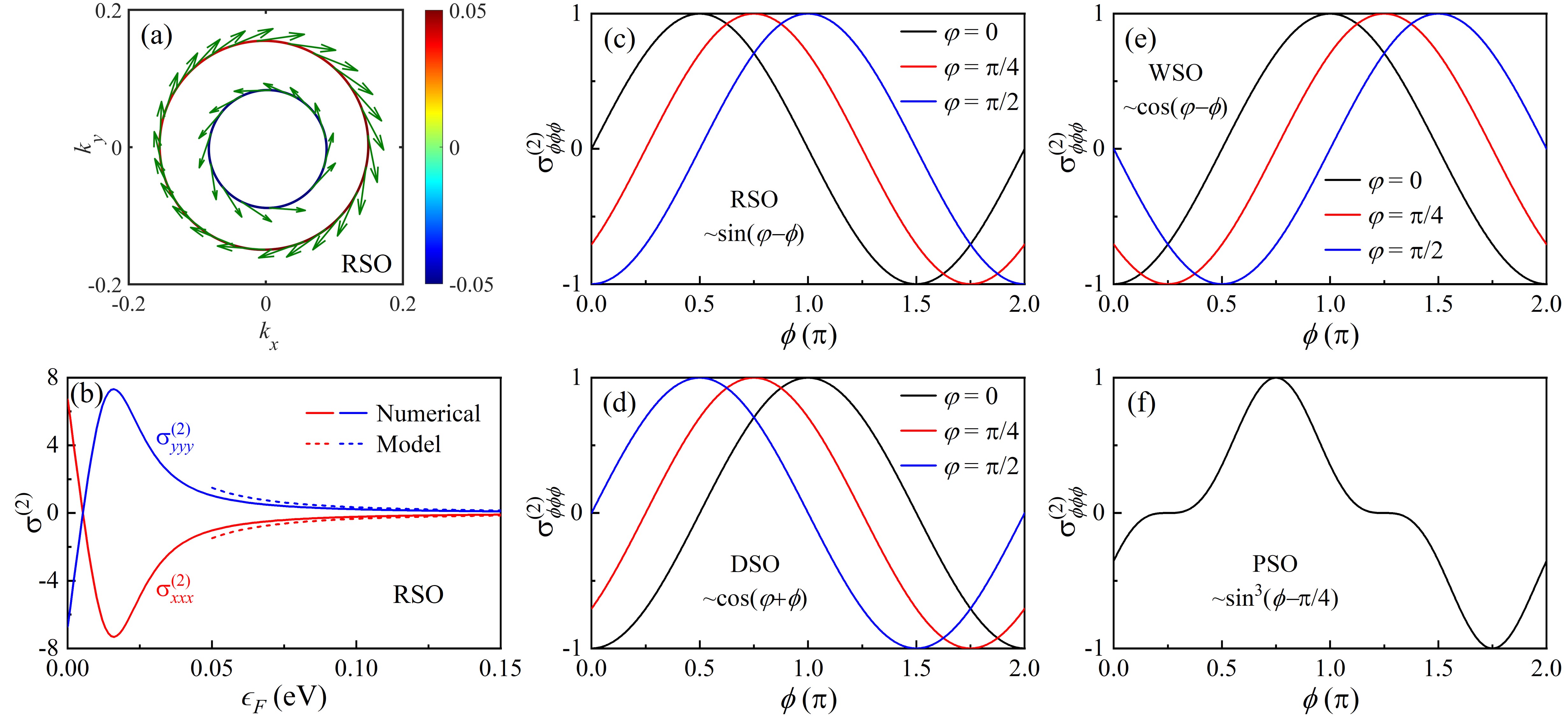}%
\caption{\label{f-1} (a) Fermi contours ($\epsilon_F=0.1$ eV) for RSO. The in-plane spin textures are indicated by the green arrows while the color map represents the out-of-plane spin polarization. $k_x$ and $k_y$ are in units of {\AA}$^{-1}$. (b) The nonlinear conductivity $\sigma^{(2)}_{xxx}$ and $\sigma^{(2)}_{yyy}$ [unit: $10^{-3}$ $e^3\hbar^3/(m^2_e|\alpha|^3)$] for RSO as a function of the Fermi energy $\epsilon_F$. The solid and dashed lines represent the numerical and model results, respectively. Normalized $\sigma^{(2)}_{\phi\phi\phi}$ for RSO (c), DSO (d), WSO (e), and PSO (f) at $\epsilon_F=0.1$ eV as a function of $\phi$ for different $\varphi$'s. In (a) and (b), $\theta$ and $\varphi$ are fixed as $\theta=\varphi=\pi/4$. In (c)-(f), $\theta$ is fixed as $\theta=\pi/4$. The other parameters are assumed to be  $m_e=0.5$ $m_0$ ($m_0$ for electron rest mass), $\alpha=\beta=\gamma=\delta=0.5$ eV {\AA}, $\Delta=0.01$ eV, and $T=50$ K in the Fermi distribution function.}
\end{figure*}

To second order in an applied electric field $\mathbf{\mathcal{E}}$, the produced current density $\mathbf{J}$ is given by\cite{callaway}
\begin{equation}\label{eq-3}
  J_a=\sigma^{(1)}_{ab}\mathcal{E}_b+\sigma^{(2)}_{abc}\mathcal{E}_{b}\mathcal{E}_{c},
\end{equation}
where $\sigma^{(1)}$ and $\sigma^{(2)}$ are the first-order linear and second-order nonlinear conductivities, respectively. The indices $a, b, c=x, y$ denote Cartesian components and a summation over repeated indices is implied. It has been proved that $\sigma^{(2)}_{abc}$ can be separated into three parts\cite{prl026301,prbL201405}
\begin{equation}\label{eq-4}
  \sigma^{(2)}_{abc}=\sigma^{\text{Drude}}_{abc}+\sigma^{\text{BCD}}_{abc}+\sigma^{\text{QM}}_{abc}.
\end{equation}
The first term $\sigma^{\text{Drude}}_{abc}$ refers to the nonlinear Drude conductivity. Under the relaxation time $\tau$ approximation, $\sigma^{\text{Drude}}_{abc}$ takes the form\cite{prr043081,prb125114,prb085419}
\begin{equation}\label{eq-5}
    \sigma^{\text{Drude}}_{abc}=-\frac{e^3\tau^2}{4\pi^2\hbar^3}\sum_n\int f_n\frac{\partial^3\epsilon_{n\mathbf{k}}}{\partial k_a \partial k_b \partial k_c}d^2k,
\end{equation}
where $f_n(\epsilon_{n\mathbf{k}}, \epsilon_F)$ is the Fermi distribution function given in terms of the eigenvalue of the $n$th band $\epsilon_{n\mathbf{k}}$ and the Fermi energy $\epsilon_F$. The second term $\sigma^{\text{BCP}}_{abc}$ represents the nonlinear conductivity induced by the Berry curvature dipole (BCD) and reads\cite{prl216806}
\begin{equation}\label{eq-6}
  \sigma^{\text{BCD}}_{abc}=-\frac{e^3\tau}{4\pi^2\hbar^2}\sum_n\int f_n(\frac{\partial\Omega^{ca}_n}{\partial k_b}+\frac{\partial\Omega^{ba}_n}{\partial k_c})d^2k,
\end{equation}
where $\Omega^{ab}_n$ is the Berry curvature given by $\Omega^{ab}_n=\partial_{k_a} A^{b}_{nn}-\partial_{k_b} A^{a}_{nn}$, with $A^{a}_{nn}$ being the Berry connection $A^{a}_{nn}=i\langle n|\partial_{k_a}|n\rangle$ ($|n\rangle$ for the $n$th Bloch state). It is clearly that $\sigma^{\text{BCD}}_{aaa}=0$ due to $\Omega^{aa}_n=0$. Thus, $\sigma^{\text{BCD}}_{abc}$ represents the nonlinear transverse (Hall) effect and does not contribute to the longitudinal nonlinear conductivity. The last term $\sigma^{\text{QM}}_{abc}$ represents the nonlinear conductivity induced by the quantum metric and is given by\cite{prl026301}
\begin{equation}\label{eq-7}
  \sigma^{\text{QM}}_{abc}=-\frac{e^3}{4\pi^2\hbar}\sum_n\int f_n[2\frac{\partial G^{bc}_n}{\partial k_a}-\frac{1}{2}(\frac{\partial G^{ca}_n}{\partial k_b}+\frac{\partial G^{ba}_n}{\partial k_c})]d^2k,
\end{equation}
where $G^{ab}_n$ is the band-normalized quantum metric or the Berry connection polarizability and reads\cite{prl277202,prl026301}
\begin{equation}\label{eq-8}
  G^{ab}_n=2\text{Re}\sum_{m\neq n}\frac{A^{a}_{nm}A^{b}_{mn}}{\epsilon_{n\mathbf{k}}-\epsilon_{m\mathbf{k}}}.
\end{equation}
For the Hamiltonian Eq. (\ref{eq-1}), the band index $n$ is replaced by the spin index $s$ in Eqs. (\ref{eq-5})-(\ref{eq-8}) and $G^{ab}_s$ can be obtained as\cite{prl166601}
\begin{equation}\label{eq-9}
  G^{ab}_s=s\frac{\partial_{k_a}\hat{\mathbf{h}}\cdot\partial_{k_b}\hat{\mathbf{h}}}{4h},
\end{equation}
with the unit vector $\hat{\mathbf{h}}=\mathbf{h}/h$. It is instructive to examine the spatial anisotropy of the nonlinear conductivity. For $\mathbf{\mathcal{E}}$ along the $\phi$ direction, with the aid of directional derivatives and the symmetry of quantum metric, one finds the anisotropic longitudinal nonlinear conductivity $\sigma^{\text{QM}}_{\phi\phi\phi}$
\begin{widetext}
\begin{equation}\label{eq-10}
\sigma^{\text{QM}}_{\phi\phi\phi}=\sigma^{\text{QM}}_{xxx}\cos^3\phi+\sigma^{\text{QM}}_{yyy}\sin^3\phi+(\sigma^{\text{QM}}_{xyy}+2\sigma^{\text{QM}}_{yxy})\sin^2\phi\cos\phi+(\sigma^{\text{QM}}_{yxx}+2\sigma^{\text{QM}}_{xxy})\sin\phi\cos^2\phi.
\end{equation}
\end{widetext}
For a given $\mathbf{\Omega(k)}$ and so $\mathbf{h(k)}$, one first calculates $G^{ab}_s$ from Eq. (\ref{eq-9}). Then, $\sigma^{\text{QM}}_{abc}$ can be obtained by plugging $G^{ab}_s$ into Eq. (\ref{eq-7}). Finally, one obtains $\sigma^{\text{QM}}_{\phi\phi\phi}$ from Eq. (\ref{eq-10}).

\section{Anisotropic nonlinear conductivity\label{sec3}}

\begin{figure*}
\includegraphics[width=0.9\textwidth]{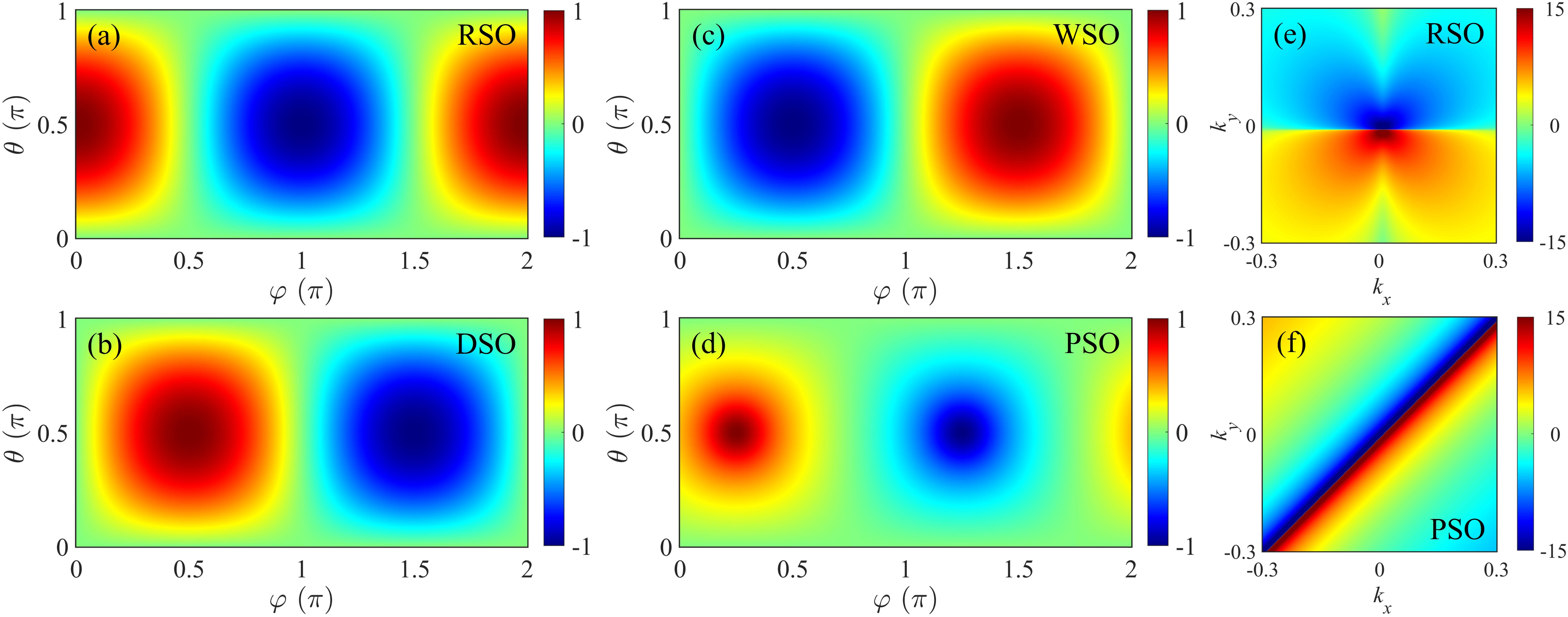}%
\caption{\label{f-2} Normalized $\sigma^{(2)}_{yyy}$ at $\epsilon_F=0.1$ eV as a function of ($\theta$, $\varphi$) for RSO (a), DSO (b), WSO (c), and PSO (d). The $\mathbf{k}$-resolved $\partial G^{yy}_+/\partial k_y$ at $\theta=\varphi=\pi/4$ for RSO (e) and PSO (f). In (e) and (f), the color map stands for
$\ln|\partial G^{yy}_+/\partial k_y|\text{sgn}(\partial G^{yy}_+/\partial k_y)$ and $k_x$, $k_y$ are in units of {\AA}$^{-1}$. The other parameters are assumed to be $m_e=0.5$ $m_0$, $\alpha=\beta=\gamma=\delta=0.5$ eV {\AA}, $\Delta=0.01$ eV, and $T=50$ K in the Fermi distribution function.}
\end{figure*}

In this section, we investigate the anisotropic nonlinear transport by examining the magnetically tunable $\sigma^{(2)}_{\phi\phi\phi}$ based on the Hamiltonian model Eq. (\ref{eq-1}). 

As an illustration, we first examine the nonlinear transport for RSO. Figure \ref{f-1}(a) shows the Fermi contours for RSO at $\epsilon_F=0.1$ eV and $\theta=\varphi=\pi/4$ and the distorted chiral spin textures can be clearly seen, indicative of band asymmetry along $k_x$ and $k_y$ directions. Figure \ref{f-1}(b) shows the numerically calculated $\sigma^{(2)}_{xxx}$ and $\sigma^{(2)}_{yyy}$ at $\theta=\varphi=\pi/4$ as a function of $\epsilon_F$. One observes that $\sigma^{(2)}_{xxx}$ and $\sigma^{(2)}_{yyy}$ are opposite and reveal slight oscillating behavior for small $\epsilon_F$. As such, $\sigma^{(2)}_{xxx}$ or $\sigma^{(2)}_{yyy}$ reaches its maximum magnitude at $\epsilon_F=16$ meV. For large $\epsilon_F$, the magnitude of $\sigma^{(2)}_{xxx}$ or $\sigma^{(2)}_{yyy}$ decreases monotonically with increasing $\epsilon_F$, which are in good accordance with model results [see the following Eqs. (\ref{eq-12}) and (\ref{eq-13})].

We then examine the anisotropic nonlinear conductivity $\sigma^{(2)}_{\phi\phi\phi}$ as shown in Fig. \ref{f-1}(c)-(f) for different SOCs. The overall trend is that $\sigma^{(2)}_{\phi\phi\phi}$ is significant anisotropic characterized by the maximum or zero nonlinear conductivity for certain $\phi$'s. Intrguingly, $\sigma^{(2)}_{\phi\phi\phi}$ for RSO, DSO or WSO is $\theta$ independent while $\sigma^{(2)}_{\phi\phi\phi}$ for PSO is both $\theta$ and $\varphi$ independent. A detailed examination reveals that $\sigma^{(2)}_{\phi\phi\phi}$ for RSO is proportional to $\sin(\varphi-\phi)$. For other SOCs, we have $\sigma^{(2)}_{\phi\phi\phi}\sim\cos(\varphi+\phi)$ for DSO, $\sigma^{(2)}_{\phi\phi\phi}\sim\cos(\varphi-\phi)$ for WSO and $\sigma^{(2)}_{\phi\phi\phi}\sim\sin^3(\phi-\pi/4)$ for PSO. Such distinct relations offer an efficient way to quantify the SOC type and shall be derived in the following.

We now turn to the magnetically tunable nonlinear conductivity. Without loss of generality, we consider the $\sigma^{(2)}_{yyy}$ component. Figure \ref{f-2}(a)-(d) show the calculated normalized $\sigma^{(2)}_{yyy}$ at $\epsilon_F=0.1$ eV as a function of ($\theta, \varphi$) for different SOCs. The overall trend is that $\sigma^{(2)}_{yyy}$ can be significantly tuned by the direction of a field, which is quite similar to that of the nonlinear Drude conductivity\cite{prb155411}. As can be seen, the magnitude of $\sigma^{(2)}_{yyy}$ reaches maximum or zero for certain ($\theta$'s, $\varphi$'s). In addition, the ($\theta, \varphi$) dependent relations for different SOCs are distinct. It is seen that $\sigma^{(2)}_{yyy}\sim\sin\theta\sin(\varphi-\pi/2)$ for RSO, $\sigma^{(2)}_{yyy}\sim\sin\theta\cos(\varphi+\pi/2)$ for DSO, $\sigma^{(2)}_{yyy}\sim\sin\theta\cos(\varphi-\pi/2)$ for WSO, and $\sigma^{(2)}_{yyy}\sim\sin\theta\sin(\varphi+\pi/4)$ for PSO. Thus, one can infer the complete ($\theta, \varphi$) dependent $\sigma^{(2)}_{\phi\phi\phi}$ from Fig. \ref{f-1}(c)-(f) and Fig. \ref{f-2}(a)-(d), that is, $\sigma^{(2)}_R\sim\sin\theta\sin(\varphi-\phi)$, $\sigma^{(2)}_D\sim\sin\theta\cos(\varphi+\phi)$, $\sigma^{(2)}_W\sim\sin\theta\cos(\varphi-\phi)$ and $\sigma^{(2)}_P\sim\sin\theta\sin(\varphi+\pi/4)\sin^3(\phi-\pi/4)$. As a comparison, the nonlinear Drude conductivity reveals the same ($\theta, \varphi$) dependent relations due to their identical symmetry restrictions\cite{prb155411}.

On the other hand, from Eq. (\ref{eq-7}), we have $\sigma^{(2)}_{yyy}\sim\sum_s\int(f_s\partial G^{yy}_s/\partial k_y)d^2k$, it is therefore helpful to examine the $\mathbf{k}$-resolved $\partial G^{yy}_s/\partial k_y$. For illustration, we consider the higher branch $\partial G^{yy}_+/\partial k_y$ at $\theta=\varphi=\pi/4$ for RSO and PSO. As shown in Fig. \ref{f-2}(e), $\partial G^{yy}_+/\partial k_y$ (logarithmic scale) for RSO is mainly contributed by the regions around the $\mathbf{k}=\mathbf{0}$ and diagonal lines $k_x=\pm k_y$. In the case of PSO shown in Fig. \ref{f-2}(f), the dominant contributions to $\partial G^{yy}_+/\partial k_y$ are from the regions close to $k_x=k_y$.

Having demonstrated the magnetic control of nonlinear transport based on numerical calculations. We now derive the analytical formulas for $\sigma_{\phi\phi\phi}^{\text{QM}}$ to explain the above observed distinct relations for different SOCs. In a zero-temperature limit, that is, $f_s(\epsilon_{\mathbf{k}s}, \epsilon_F)=\theta(\epsilon_F-\epsilon_{\mathbf{k}s})$, substitution of Eq. (\ref{eq-9}) in Eq. (\ref{eq-7}) yileds
\begin{widetext}
\begin{equation}\label{eq-11}
  \sigma^{\text{QM}}_{abc}=-\frac{e^3}{4\pi^2\hbar}\int^{2\pi}_0d\phi\int^{k_{F+}}_{k_{F-}}[2\frac{\partial G^{bc}_+}{\partial k_a}-\frac{1}{2}(\frac{\partial G^{ca}_+}{\partial k_b}+\frac{\partial G^{ba}_+}{\partial k_c})]kdk,
\end{equation}
\end{widetext}
where $k_{F\pm}$ are spin-split Fermi wave numbers. In the following derivations, we consider the weak-field or high-density regime, namely $|\alpha| k_F\gg|\Delta|$. For RSO, we obtain
\begin{equation}\label{eq-12}
\sigma_{\phi\phi\phi}^{\text{QM}}\approx\sigma^{\text{QM}}_{R}\text{sin}\theta \text{sin}(\varphi-\phi),
\end{equation}
where the prefactor $\sigma^{\text{QM}}_R$ is defined as
\begin{equation}\label{eq-13} 
\sigma^{\text{QM}}_R=-\frac{5e^3(3\hbar^2\epsilon_F+2m_e\alpha^2)\Delta}{128\pi\hbar m_e\alpha\epsilon^3_F}.
\end{equation}
For DSO, we obtain
\begin{equation}\label{eq-14}
\sigma_{\phi\phi\phi}^{\text{QM}}\approx\sigma^{\text{QM}}_{D}\sin\theta\cos(\varphi+\phi).
\end{equation}
For WSO, we find
\begin{equation}\label{eq-15}
\sigma_{\phi\phi\phi}^{\text{QM}}\approx\sigma^{\text{QM}}_{W}\sin\theta\cos(\varphi-\phi).
\end{equation}
In Eqs. (\ref{eq-14}) and (\ref{eq-15}), the prefactors $\sigma^{\text{QM}}_{D}$ and $\sigma^{\text{QM}}_{W}$ can be obtained by replacing $\alpha$ by $\beta$ and $\gamma$ in Eq. (\ref{eq-13}), respectively. For PSO, we find
\begin{equation}\label{eq-16}
\sigma_{\phi\phi\phi}^{\text{QM}}\approx\sigma^{\text{QM}}_P\sin\theta\sin(\varphi+\frac{\pi}{4})\sin^3(\phi-\frac{\pi}{4}),
\end{equation}
where the prefactor $\sigma^{\text{QM}}_P$ is given by Eq. (\ref{aq-27}) in Appendix \ref{secA}. Details of the derivation of Eqs. (\ref{eq-12})-(\ref{eq-16}) are presented in Appendix \ref{secA}. We find that Eqs. (\ref{eq-12})-(\ref{eq-16}) (as listed in Table \ref{table1}) are in exact accordance with numerical results shown in Figs. \ref{f-1} and \ref{f-2}.

It is instructive to compare the quantum-metric-induced nonlinear conductivity $\sigma_{\phi\phi\phi}^{\text{QM}}$ with nonlinear Drude conductivity  $\sigma_{\phi\phi\phi}^{\text{Drude}}$ investigated previously\cite{prb155411}. As listed in Table \ref{table1}, there is the same ($\theta, \varphi$) dependent relations for the same SOC in both due to their identical symmetry restrictions. However, for RSO, DSO and WSO, the signs of $\sigma_{\phi\phi\phi}^{\text{QM}}$ and $\sigma_{\phi\phi\phi}^{\text{Drude}}$ for the same SOC are opposite, which results in the compensated contribution to the total nonlinear conductivity. For example, $\sigma^{\text{Drude}}_R=e^3\tau^2\alpha\Delta/(4\pi\hbar^3\epsilon_F)$\cite{prb155411} and $\sigma^{\text{QM}}_R$ [Eq. (\ref{eq-13})] have opposite signs but reveal distinct $\epsilon_F$ dependency relations. In the case of PSO, $\sigma_{\phi\phi\phi}^{\text{Drude}}$ and $\sigma_{\phi\phi\phi}^{\text{QM}}$ have the same sign but distinct $\phi$ dependency relations: the former is $\sin(\phi-\pi/4)$ dependent and the latter is $\sin^3(\phi-\pi/4)$ dependent. As such, it is feasible to seperate different contributions from the sign of nonlinear conductivity or its $\phi$ dependent relations.

\section{Discussion and conclusions\label{sec4}}

In this work, we study the magnetically tunable nonlinear transport based on the Hamiltonian model. For realistic systems, it is promising to consider the 2D material with SOC on top of a ferromagnetic or antiferromagnetic insulator. For example, an Ag$_2$Te monolayer with RSO on top of the Cr$_2$O$_3$ is one of the promising candidates\cite{prb155411,npj172}. An alternative way is by applying an external magnetic field on 2D materials. The density functional theory calculations on realistic systems will be provided in the future study. Second, the similar approach can be readily generalized to three-dimensional systems. Lastly, since the polarity of the nonlinear conductivity is locked to the magnetic order parameters such as magnetization and N\'{e}el vector, measuring the magnetic control of nonlinear conductivity offers an efficient way to detect those magnetic order parameters, which is beneficial for future devices. 

In summary, using the Boltzmann transport theory, we have studied the magnetic control of nonlinear transport induced by the quantum metric in two dimensions. Our results show that the nonlinear conductivity can be significantly tuned by the direction of a field and reveals strong spatial anisotropy. We further derive the analytical formulas of the nonlinear conductivity for different SOCs, which are in exact agreement with numerical results. Our work establishes a fundamental strategy in exploring the nonlinear transport physics due to the \emph{intrinsic} quantum geometry.

\begin{center}
{\bf ACKNOWLEDGMENTS}
\end{center}

This research was supported by the National Natural Science Foundation of China (Grant No. 12274102).

\begin{center}
{\bf DATA AVAILABILITY}
\end{center}

The data that support the findings of this article are not publicly available upon publication because it is not technically feasible and/or the cost of preparing, depositing, and hosting the data would be prohibitive within the terms of this research project. The data are available from the authors upon reasonable request.

\begin{widetext}

\appendix
\section{Derivation of Eqs. (\ref{eq-12})-(\ref{eq-16})\label{secA}}

For RSO, we have $\mathbf{\Omega(k)}=\alpha(-k_y, k_x, 0)$ and $\mathbf{h(k)}=(-\alpha k_y-\Delta\hat{m}_x, \alpha k_x-\Delta\hat{m}_y, -\Delta\hat{m}_z)$. It follows from Eq. (\ref{eq-9}) that
\begin{equation}
\label{aq-1}
G^{xx}_s=s\frac{\alpha^2(h^2_x+h^2_z)}{4h^5}, G^{yy}_s=s\frac{\alpha^2(h^2_y+h^2_z)}{4h^5}, G^{xy}_s=s\frac{\alpha^2h_xh_y}{4h^5}.
\end{equation}
In zero order, the Fermi wave number $k_{Fs}$ can be calculated as $k_{Fs}\approx\sqrt{k_0^2+k_R^2}-sk_R$ with $k_0=\sqrt{2m_e\epsilon_F/\hbar^2}$ and $k_R=m_e|\alpha|/\hbar^2$. We substitute in Eq. (\ref{eq-11}) and find that
\begin{equation}\label{aq-2}
\begin{aligned}
&\sigma^{\text{QM}}_{xxx}=-\frac{e^3}{4\pi^2\hbar}\sum_s\int f_s\frac{\partial G^{xx}_s}{\partial k_x}d^2k=-\frac{5\alpha^3e^3}{16\pi^2\hbar}\int^{2\pi}_0d\phi\int^{k_{F+}}_{k_{F-}}(\frac{h^3_y}{h^7}-\frac{h_y}{h^5})kdk,\\
&\sigma^{\text{QM}}_{yyy}=-\frac{e^3}{4\pi^2\hbar}\sum_s\int f_s\frac{\partial G^{yy}_s}{\partial k_y}d^2k=-\frac{5\alpha^3e^3}{16\pi^2\hbar}\int^{2\pi}_0d\phi\int^{k_{F+}}_{k_{F-}}(\frac{h_x}{h^5}-\frac{h^3_x}{h^7})kdk,\\
&\sigma^{\text{QM}}_{yxx}=-\frac{e^3}{4\pi^2\hbar}\sum_s\int f_s(2\frac{\partial G^{xx}_s}{\partial k_y}-\frac{\partial G^{xy}_s}{\partial k_x})d^2k\approx-\frac{5\alpha^3e^3}{16\pi^2\hbar}\int^{2\pi}_0d\phi\int^{k_{F+}}_{k_{F-}}\frac{h^3_x}{h^7}kdk,\\
&\sigma^{\text{QM}}_{xxy}=-\frac{e^3}{4\pi^2\hbar}\sum_s\int f_s(\frac{3}{2}\frac{\partial G^{xy}_s}{\partial k_x}-\frac{1}{2}\frac{\partial G^{xx}_s}{\partial k_y})d^2k\approx\frac{5\alpha^3e^3}{16\pi^2\hbar}\int^{2\pi}_0d\phi\int^{k_{F+}}_{k_{F-}}(\frac{h_x}{h^5}-\frac{h^3_x}{h^7})kdk,\\
&\sigma^{\text{QM}}_{yxy}=-\frac{e^3}{4\pi^2\hbar}\sum_s\int f_s(\frac{3}{2}\frac{\partial G^{xy}_s}{\partial k_y}-\frac{1}{2}\frac{\partial G^{yy}_s}{\partial k_x})d^2k\approx\frac{5\alpha^3e^3}{16\pi^2\hbar}\int^{2\pi}_0d\phi\int^{k_{F+}}_{k_{F-}}(\frac{h^3_y}{h^7}-\frac{h_y}{h^5})kdk,\\
&\sigma^{\text{QM}}_{xyy}=-\frac{e^3}{4\pi^2\hbar}\sum_s\int f_s(2\frac{\partial G^{yy}_s}{\partial k_x}-\frac{\partial G^{yx}_s}{\partial k_y})d^2k\approx\frac{5\alpha^3e^3}{16\pi^2\hbar}\int^{2\pi}_0d\phi\int^{k_{F+}}_{k_{F-}}\frac{h^3_y}{h^7}kdk.\\
\end{aligned}
\end{equation}
To first order in $\Delta$, the integrals in Eq. (\ref{aq-2}) can be calculated as
\begin{equation}\label{aq-3}
\begin{aligned}
&\int^{2\pi}_0d\phi\int^{k_{F+}}_{k_{F-}}\frac{h^3_y}{h^7}kdk\approx\int^{2\pi}_0d\phi\int^{k_{F+}}_{k_{F-}}(\alpha^3k^3_x-3\alpha^2k^2_x\Delta\hat{m}_y)[\frac{1}{|\alpha|^7k^6}+\frac{7\alpha\Delta}{|\alpha|^9k^7}(\hat{m}_y\cos\phi-\hat{m}_x\sin\phi)]dk\\
&\approx\int^{2\pi}_0d\phi\int^{k_{F+}}_{k_{F-}}[\frac{\cos^3\phi}{\alpha^3|\alpha|k^3}+\frac{7\Delta\cos^3\phi}{|\alpha|^5k^4}(\hat{m}_y\cos\phi-\hat{m}_x\sin\phi)-\frac{3\Delta\hat{m}_y\cos^2\phi}{|\alpha|^5k^4}]dk\\
&=\int^{2\pi}_0[\frac{\cos^3\phi}{\alpha^3|\alpha|}(\frac{1}{2k^2_{F-}}-\frac{1}{2k^2_{F+}})+\frac{\Delta\hat{m}_y(7\cos^4\phi-3\cos^2\phi)-7\Delta\hat{m}_x\cos^3\phi\sin\phi}{|\alpha|^5}(\frac{1}{3k^3_{F-}}-\frac{1}{3k^3_{F+}})]d\phi\\
&\approx\int^{2\pi}_0[\frac{\cos^3\phi}{\alpha^3|\alpha|}(-\frac{2k_R\sqrt{k^2_0+k^2_R}}{k^4_0})+\frac{\Delta\hat{m}_y(7\cos^4\phi-3\cos^2\phi)-7\Delta\hat{m}_x\cos^3\phi\sin\phi}{|\alpha|^5}(-\frac{2k_R(3k^2_0+4k^2_R)}{3k^6_0})]d\phi\\
&=-\frac{2k_R(3k^2_0+4k^2_R)}{3k^6_0}\int^{2\pi}_0\frac{\Delta\hat{m}_y(7\cos^4\phi-3\cos^2\phi)}{|\alpha|^5}d\phi\\
&=-\frac{3\pi(3\hbar^2\epsilon_F+2m_e\alpha^2)\Delta}{8m_e\alpha^4\epsilon^3_F}\sin\theta\sin\varphi,
\end{aligned}
\end{equation}
and
\begin{equation}\label{aq-4}
\begin{aligned}
&\int^{2\pi}_0d\phi\int^{k_{F+}}_{k_{F-}}\frac{h_y}{h^5}kdk\approx\int^{2\pi}_0d\phi\int^{k_{F+}}_{k_{F-}}(\alpha k_x-\Delta\hat{m}_y)[\frac{1}{|\alpha|^5k^4}+\frac{5\alpha\Delta}{|\alpha|^7k^5}(\hat{m}_y\cos\phi-\hat{m}_x\sin\phi)]dk\\
&\approx\int^{2\pi}_0d\phi\int^{k_{F+}}_{k_{F-}}[\frac{\alpha\cos\phi}{|\alpha|^5k^3}+\frac{5\Delta\cos\phi}{|\alpha|^5k^4}(\hat{m}_y\cos\phi-\hat{m}_x\sin\phi)-\frac{\Delta\hat{m}_y}{|\alpha|^5k^4}]dk\\
&=\int^{2\pi}_0[\frac{\alpha\cos\phi}{|\alpha|^5}(\frac{1}{2k^2_{F-}}-\frac{1}{2k^2_{F+}})+\frac{\Delta}{|\alpha|^5}(5\hat{m}_y\cos^2\phi-5\hat{m}_x\cos\phi\sin\phi-\hat{m}_y)(\frac{1}{3k^3_{F-}}-\frac{1}{3k^3_{F+}})]d\phi\\
&\approx\int^{2\pi}_0[\frac{\alpha\cos\phi}{|\alpha|^5}(-\frac{2k_R\sqrt{k^2_0+k^2_R}}{k^4_0})+\frac{\Delta}{|\alpha|^5}(5\hat{m}_y\cos^2\phi-5\hat{m}_x\cos\phi\sin\phi-\hat{m}_y)(-\frac{2k_R(3k^2_0+4k^2_R)}{3k^6_0})]d\phi\\
&=-\frac{2k_R(3k^2_0+4k^2_R)}{3k^6_0}\int^{2\pi}_0[\frac{\Delta\hat{m}_y}{|\alpha|^5}(5\cos^2\phi-1)]d\phi\\
&=-\frac{\pi(3\hbar^2\epsilon_F+2m_e\alpha^2)\Delta}{2m_e\alpha^4\epsilon^3_F}\sin\theta\sin\varphi,
\end{aligned}
\end{equation}
and
\begin{equation}\label{aq-5}
\begin{aligned}
&\int^{2\pi}_0d\phi\int^{k_{F+}}_{k_{F-}}\frac{h_x}{h^5}kdk\approx-\int^{2\pi}_0d\phi\int^{k_{F+}}_{k_{F-}}(\alpha k_y+\Delta\hat{m}_x)[\frac{1}{|\alpha|^5k^4}+\frac{5\alpha\Delta}{|\alpha|^7k^5}(\hat{m}_y\cos\phi-\hat{m}_x\sin\phi)]dk\\
&\approx-\int^{2\pi}_0d\phi\int^{k_{F+}}_{k_{F-}}[\frac{\alpha\sin\phi}{|\alpha|^5k^3}+\frac{5\Delta\sin\phi}{|\alpha|^5k^4}(\hat{m}_y\cos\phi-\hat{m}_x\sin\phi)+\frac{\Delta\hat{m}_x}{|\alpha|^5k^4}]dk\\
&=-\int^{2\pi}_0[\frac{\alpha\sin\phi}{|\alpha|^5}(\frac{1}{2k^2_{F-}}-\frac{1}{2k^2_{F+}})+\frac{\Delta}{|\alpha|^5}(5\hat{m}_y\cos\phi\sin\phi-5\hat{m}_x\sin^2\phi+\hat{m}_x)(\frac{1}{3k^3_{F-}}-\frac{1}{3k^3_{F+}})]d\phi\\
&\approx-\frac{2k_R(3k^2_0+4k^2_R)}{3k^6_0}\int^{2\pi}_0[\frac{\Delta\hat{m}_x}{|\alpha|^5}(5\sin^2\phi-1)]d\phi\\
&=-\frac{\pi(3\hbar^2\epsilon_F+2m_e\alpha^2)\Delta}{2m_e\alpha^4\epsilon^3_F}\sin\theta\cos\varphi,
\end{aligned}
\end{equation}
and
\begin{equation}\label{aq-6}
\begin{aligned}
&\int^{2\pi}_0d\phi\int^{k_{F+}}_{k_{F-}}\frac{h^3_x}{h^7}kdk\approx-\int^{2\pi}_0d\phi\int^{k_{F+}}_{k_{F-}}(\alpha^3k^3_y+3\alpha^2k^2_y\Delta\hat{m}_x)[\frac{1}{|\alpha|^7k^6}+\frac{7\alpha\Delta}{|\alpha|^9k^7}(\hat{m}_y\cos\phi-\hat{m}_x\sin\phi)]dk\\
&\approx-\int^{2\pi}_0d\phi\int^{k_{F+}}_{k_{F-}}[\frac{\sin^3\phi}{\alpha^3|\alpha|k^3}+\frac{7\Delta\sin^3\phi}{|\alpha|^5k^4}(\hat{m}_y\cos\phi-\hat{m}_x\sin\phi)+\frac{3\Delta\hat{m}_x\sin^2\phi}{|\alpha|^5k^4}]dk\\
&\approx-\frac{2k_R(3k^2_0+4k^2_R)}{3k^6_0}\int^{2\pi}_0\frac{\Delta\hat{m}_x(7\sin^4\phi-3\sin^2\phi)}{|\alpha|^5}d\phi\\
&=-\frac{3\pi(3\hbar^2\epsilon_F+2m_e\alpha^2)\Delta}{8m_e\alpha^4\epsilon^3_F}\sin\theta\cos\varphi.
\end{aligned}
\end{equation}
Substitution of Eqs. (\ref{aq-3})-(\ref{aq-6}) in Eq. (\ref{aq-2}) yields
\begin{equation}\label{aq-7}
\begin{aligned}
&\sigma^{\text{QM}}_{xxx}=\sigma^{\text{QM}}_R\sin\theta\sin\varphi, \sigma^{\text{QM}}_{yyy}=-\sigma^{\text{QM}}_R\sin\theta\cos\varphi, \sigma^{\text{QM}}_{yxx}=-3\sigma^{\text{QM}}_R\sin\theta\cos\varphi,\\
&\sigma^{\text{QM}}_{xxy}=\sigma^{\text{QM}}_R\sin\theta\cos\varphi, \sigma^{\text{QM}}_{yxy}=-\sigma^{\text{QM}}_R\sin\theta\sin\varphi, \sigma^{\text{QM}}_{xyy}=3\sigma^{\text{QM}}_R\sin\theta\sin\varphi,\\
&\sigma^{\text{QM}}_R\equiv-\frac{5e^3(3\hbar^2\epsilon_F+2m_e\alpha^2)\Delta}{128\pi\hbar m_e\alpha\epsilon^3_F}.
\end{aligned}
\end{equation}
We substitute in Eq. (\ref{eq-10}) and find that
\begin{equation}\label{aq-8}
\begin{aligned}
\sigma^{\text{QM}}_{\phi\phi\phi}&=\sigma^{\text{QM}}_{xxx}\cos^3\phi+\sigma^{\text{QM}}_{yyy}\sin^3\phi+(\sigma^{\text{QM}}_{xyy}+2\sigma^{\text{QM}}_{yxy})\sin^2\phi\cos\phi+(\sigma^{\text{QM}}_{yxx}+2\sigma^{\text{QM}}_{xxy})\sin\phi\cos^2\phi\\
&=\sigma^{\text{QM}}_{xxx}\cos^3\phi+\sigma^{\text{QM}}_{yyy}\sin^3\phi+(3\sigma^{\text{QM}}_{xxx}-2\sigma^{\text{QM}}_{xxx})\sin^2\phi\cos\phi+(3\sigma^{\text{QM}}_{yyy}-2\sigma^{\text{QM}}_{yyy})\sin\phi\cos^2\phi\\
&=\sigma^{\text{QM}}_{xxx}\cos\phi+\sigma^{\text{QM}}_{yyy}\sin\phi=\sigma^{\text{QM}}_R\sin\theta\sin(\varphi-\phi).
\end{aligned}
\end{equation}
The above procedures to derive $\sigma^{\text{QM}}_{\phi\phi\phi}$ can be readily generalized to other SOCs. For DSO, we have $\mathbf{h(k)}=(\beta k_x-\Delta\hat{m}_x, -\beta k_y-\Delta\hat{m}_y, -\Delta\hat{m}_z)$ and
\begin{equation}\label{aq-9}
G^{xx}_s=s\frac{\beta^2(h^2_y+h^2_z)}{4h^5}, G^{yy}_s=s\frac{\beta^2(h^2_x+h^2_z)}{4h^5}, G^{xy}_s=s\frac{\beta^2h_xh_y}{4h^5}.
\end{equation}
We substitute in Eq. (\ref{eq-11}) and find that
\begin{equation}\label{aq-10}
\begin{aligned}
&\sigma^{\text{QM}}_{xxx}=-\frac{e^3}{4\pi^2\hbar}\sum_s\int f_s\frac{\partial G^{xx}_s}{\partial k_x}d^2k=-\frac{5\beta^3e^3}{16\pi^2\hbar}\int^{2\pi}_0d\phi\int^{k_{F+}}_{k_{F-}}(\frac{h^3_x}{h^7}-\frac{h_x}{h^5})kdk,\\
&\sigma^{\text{QM}}_{yyy}=-\frac{e^3}{4\pi^2\hbar}\sum_s\int f_s\frac{\partial G^{yy}_s}{\partial k_y}d^2k=-\frac{5\beta^3e^3}{16\pi^2\hbar}\int^{2\pi}_0d\phi\int^{k_{F+}}_{k_{F-}}(\frac{h_y}{h^5}-\frac{h^3_y}{h^7})kdk,\\
&\sigma^{\text{QM}}_{yxx}=-\frac{e^3}{4\pi^2\hbar}\sum_s\int f_s(2\frac{\partial G^{xx}_s}{\partial k_y}-\frac{\partial G^{xy}_s}{\partial k_x})d^2k\approx-\frac{5\beta^3e^3}{16\pi^2\hbar}\int^{2\pi}_0d\phi\int^{k_{F+}}_{k_{F-}}\frac{h^3_y}{h^7}kdk,\\
&\sigma^{\text{QM}}_{xxy}=-\frac{e^3}{4\pi^2\hbar}\sum_s\int f_s(\frac{3}{2}\frac{\partial G^{xy}_s}{\partial k_x}-\frac{1}{2}\frac{\partial G^{xx}_s}{\partial k_y})d^2k\approx\frac{5\beta^3e^3}{16\pi^2\hbar}\int^{2\pi}_0d\phi\int^{k_{F+}}_{k_{F-}}(\frac{h_y}{h^5}-\frac{h^3_y}{h^7})kdk,\\
&\sigma^{\text{QM}}_{yxy}=-\frac{e^3}{4\pi^2\hbar}\sum_s\int f_s(\frac{3}{2}\frac{\partial G^{xy}_s}{\partial k_y}-\frac{1}{2}\frac{\partial G^{yy}_s}{\partial k_x})d^2k\approx\frac{5\beta^3e^3}{16\pi^2\hbar}\int^{2\pi}_0d\phi\int^{k_{F+}}_{k_{F-}}(\frac{h^3_x}{h^7}-\frac{h_x}{h^5})kdk,\\
&\sigma^{\text{QM}}_{xyy}=-\frac{e^3}{4\pi^2\hbar}\sum_s\int f_s(2\frac{\partial G^{yy}_s}{\partial k_x}-\frac{\partial G^{yx}_s}{\partial k_y})d^2k\approx\frac{5\beta^3e^3}{16\pi^2\hbar}\int^{2\pi}_0d\phi\int^{k_{F+}}_{k_{F-}}\frac{h^3_x}{h^7}kdk.\\
\end{aligned}
\end{equation}
To first order in $\Delta$, we have
\begin{equation}\label{aq-11}
\begin{aligned}
&\int^{2\pi}_0d\phi\int^{k_{F+}}_{k_{F-}}\frac{h^3_x}{h^7}kdk\approx\int^{2\pi}_0d\phi\int^{k_{F+}}_{k_{F-}}(\beta^3k^3_x-3\beta^2k^2_x\Delta\hat{m}_x)[\frac{1}{|\beta|^7k^6}+\frac{7\beta\Delta}{|\beta|^9k^7}(\hat{m}_x\cos\phi-\hat{m}_y\sin\phi)]dk\\
&\approx\int^{2\pi}_0d\phi\int^{k_{F+}}_{k_{F-}}[\frac{\cos^3\phi}{\beta^3|\beta|k^3}+\frac{7\Delta\cos^3\phi}{|\beta|^5k^4}(\hat{m}_x\cos\phi-\hat{m}_y\sin\phi)-\frac{3\Delta\hat{m}_x\cos^2\phi}{|\beta|^5k^4}]dk\\
&=\int^{2\pi}_0[\frac{\cos^3\phi}{\beta^3|\beta|}(\frac{1}{2k^2_{F-}}-\frac{1}{2k^2_{F+}})+\frac{\Delta\hat{m}_x(7\cos^4\phi-3\cos^2\phi)-7\Delta\hat{m}_y\cos^3\phi\sin\phi}{|\beta|^5}(\frac{1}{3k^3_{F-}}-\frac{1}{3k^3_{F+}})]d\phi\\
&\approx-\frac{2k_D(3k^2_0+4k^2_D)}{3k^6_0}\int^{2\pi}_0\frac{\Delta\hat{m}_x(7\cos^4\phi-3\cos^2\phi)}{|\beta|^5}d\phi\\
&=-\frac{3\pi(3\hbar^2\epsilon_F+2m_e\beta^2)\Delta}{8m_e\beta^4\epsilon^3_F}\sin\theta\cos\varphi,
\end{aligned}
\end{equation}
and
\begin{equation}\label{aq-12}
\begin{aligned}
&\int^{2\pi}_0d\phi\int^{k_{F+}}_{k_{F-}}\frac{h_x}{h^5}kdk\approx\int^{2\pi}_0d\phi\int^{k_{F+}}_{k_{F-}}(\beta k_x-\Delta\hat{m}_x)[\frac{1}{|\beta|^5k^4}+\frac{5\beta\Delta}{|\beta|^7k^5}(\hat{m}_x\cos\phi-\hat{m}_y\sin\phi)]dk\\
&\approx\int^{2\pi}_0d\phi\int^{k_{F+}}_{k_{F-}}[\frac{\beta\cos\phi}{|\beta|^5k^3}+\frac{5\Delta\cos\phi}{|\beta|^5k^4}(\hat{m}_x\cos\phi-\hat{m}_y\sin\phi)-\frac{\Delta\hat{m}_x}{|\beta|^5k^4}]dk\\
&=\int^{2\pi}_0[\frac{\beta\cos\phi}{|\beta|^5}(\frac{1}{2k^2_{F-}}-\frac{1}{2k^2_{F+}})+\frac{\Delta}{|\beta|^5}(5\hat{m}_x\cos^2\phi-5\hat{m}_y\cos\phi\sin\phi-\hat{m}_x)(\frac{1}{3k^3_{F-}}-\frac{1}{3k^3_{F+}})]d\phi\\
&\approx-\frac{2k_D(3k^2_0+4k^2_D)}{3k^6_0}\int^{2\pi}_0[\frac{\Delta\hat{m}_x}{|\beta|^5}(5\cos^2\phi-1)]d\phi\\
&=-\frac{\pi(3\hbar^2\epsilon_F+2m_e\beta^2)\Delta}{2m_e\beta^4\epsilon^3_F}\sin\theta\cos\varphi,
\end{aligned}
\end{equation}
and
\begin{equation}\label{aq-13}
\begin{aligned}
&\int^{2\pi}_0d\phi\int^{k_{F+}}_{k_{F-}}\frac{h_y}{h^5}kdk\approx\int^{2\pi}_0d\phi\int^{k_{F+}}_{k_{F-}}(-\beta k_y-\Delta\hat{m}_y)[\frac{1}{|\beta|^5k^4}+\frac{5\beta\Delta}{|\beta|^7k^5}(\hat{m}_x\cos\phi-\hat{m}_y\sin\phi)]dk\\
&\approx-\int^{2\pi}_0d\phi\int^{k_{F+}}_{k_{F-}}[\frac{\beta\sin\phi}{|\beta|^5k^3}+\frac{5\Delta\sin\phi}{|\beta|^5k^4}(\hat{m}_x\cos\phi-\hat{m}_y\sin\phi)+\frac{\Delta\hat{m}_y}{|\beta|^5k^4}]dk\\
&=-\int^{2\pi}_0[\frac{\beta\sin\phi}{|\beta|^5}(\frac{1}{2k^2_{F-}}-\frac{1}{2k^2_{F+}})+\frac{\Delta}{|\beta|^5}(5\hat{m}_x\cos\phi\sin\phi-5\hat{m}_y\sin^2\phi+\hat{m}_y)(\frac{1}{3k^3_{F-}}-\frac{1}{3k^3_{F+}})]d\phi\\
&\approx-\frac{2k_D(3k^2_0+4k^2_D)}{3k^6_0}\int^{2\pi}_0[\frac{\Delta\hat{m}_y}{|\beta|^5}(5\sin^2\phi-1)]d\phi\\
&=-\frac{\pi(3\hbar^2\epsilon_F+2m_e\beta^2)\Delta}{2m_e\beta^4\epsilon^3_F}\sin\theta\sin\varphi,
\end{aligned}
\end{equation}
and
\begin{equation}\label{aq-14}
\begin{aligned}
&\int^{2\pi}_0d\phi\int^{k_{F+}}_{k_{F-}}\frac{h^3_y}{h^7}kdk\approx-\int^{2\pi}_0d\phi\int^{k_{F+}}_{k_{F-}}(\beta^3k^3_y+3\beta^2k^2_y\Delta\hat{m}_y)[\frac{1}{|\beta|^7k^6}+\frac{7\beta\Delta}{|\beta|^9k^7}(\hat{m}_x\cos\phi-\hat{m}_y\sin\phi)]dk\\
&\approx-\int^{2\pi}_0d\phi\int^{k_{F+}}_{k_{F-}}[\frac{\sin^3\phi}{\beta^3|\beta|k^3}+\frac{7\Delta\sin^3\phi}{|\beta|^5k^4}(\hat{m}_x\cos\phi-\hat{m}_y\sin\phi)+\frac{3\Delta\hat{m}_y\sin^2\phi}{|\beta|^5k^4}]dk\\
&\approx-\frac{2k_D(3k^2_0+4k^2_D)}{3k^6_0}\int^{2\pi}_0\frac{\Delta\hat{m}_y(7\sin^4\phi-3\sin^2\phi)}{|\beta|^5}d\phi\\
&=-\frac{3\pi(3\hbar^2\epsilon_F+2m_e\beta^2)\Delta}{8m_e\beta^4\epsilon^3_F}\sin\theta\sin\varphi.
\end{aligned}
\end{equation}
with $k_0=\sqrt{2m_e\epsilon_F/\hbar^2}$ and $k_D=m_e|\beta|/\hbar^2$. Substitution of Eqs. (\ref{aq-11})-(\ref{aq-14}) in Eq. (\ref{aq-10}) yields
\begin{equation}\label{aq-15}
\begin{aligned}
&\sigma^{\text{QM}}_{xxx}=\sigma^{\text{QM}}_D\sin\theta\cos\varphi, \sigma^{\text{QM}}_{yyy}=-\sigma^{\text{QM}}_D\sin\theta\sin\varphi, \sigma^{\text{QM}}_{yxx}=-3\sigma^{\text{QM}}_D\sin\theta\sin\varphi,\\
&\sigma^{\text{QM}}_{xxy}=\sigma^{\text{QM}}_D\sin\theta\sin\varphi, \sigma^{\text{QM}}_{yxy}=-\sigma^{\text{QM}}_D\sin\theta\cos\varphi, \sigma^{\text{QM}}_{xyy}=3\sigma^{\text{QM}}_D\sin\theta\cos\varphi,\\
&\sigma^{\text{QM}}_D\equiv-\frac{5e^3(3\hbar^2\epsilon_F+2m_e\beta^2)\Delta}{128\pi\hbar m_e\beta\epsilon^3_F}.
\end{aligned}
\end{equation}
We substitute in Eq. (\ref{eq-10}) and find that
\begin{equation}\label{aq-16}
\begin{aligned}
\sigma^{\text{QM}}_{\phi\phi\phi}&=\sigma^{\text{QM}}_{xxx}\cos^3\phi+\sigma^{\text{QM}}_{yyy}\sin^3\phi+(\sigma^{\text{QM}}_{xyy}+2\sigma^{\text{QM}}_{yxy})\sin^2\phi\cos\phi+(\sigma^{\text{QM}}_{yxx}+2\sigma^{\text{QM}}_{xxy})\sin\phi\cos^2\phi\\
&=\sigma^{\text{QM}}_{xxx}\cos^3\phi+\sigma^{\text{QM}}_{yyy}\sin^3\phi+(3\sigma^{\text{QM}}_{xxx}-2\sigma^{\text{QM}}_{xxx})\sin^2\phi\cos\phi+(3\sigma^{\text{QM}}_{yyy}-2\sigma^{\text{QM}}_{yyy})\sin\phi\cos^2\phi\\
&=\sigma^{\text{QM}}_{xxx}\cos\phi+\sigma^{\text{QM}}_{yyy}\sin\phi=\sigma^{\text{QM}}_D\sin\theta\cos(\varphi+\phi).
\end{aligned}
\end{equation}
For WSO, we have $\mathbf{h(k)}=(\gamma k_x-\Delta\hat{m}_x, \gamma k_y-\Delta\hat{m}_y, -\Delta\hat{m}_z)$ and
\begin{equation}\label{aq-17}
G^{xx}_s=s\frac{\gamma^2(h^2_y+h^2_z)}{4h^5}, G^{yy}_s=s\frac{\gamma^2(h^2_x+h^2_z)}{4h^5}, G^{xy}_s=-s\frac{\gamma^2h_xh_y}{4h^5}.
\end{equation}
We substitute in Eq. (\ref{eq-11}) and find that
\begin{equation}\label{aq-18}
\begin{aligned}
&\sigma^{\text{QM}}_{xxx}=-\frac{e^3}{4\pi^2\hbar}\sum_s\int f_s\frac{\partial G^{xx}_s}{\partial k_x}d^2k=-\frac{5\gamma^3e^3}{16\pi^2\hbar}\int^{2\pi}_0d\phi\int^{k_{F+}}_{k_{F-}}(\frac{h^3_x}{h^7}-\frac{h_x}{h^5})kdk,\\
&\sigma^{\text{QM}}_{yyy}=-\frac{e^3}{4\pi^2\hbar}\sum_s\int f_s\frac{\partial G^{yy}_s}{\partial k_y}d^2k=-\frac{5\gamma^3e^3}{16\pi^2\hbar}\int^{2\pi}_0d\phi\int^{k_{F+}}_{k_{F-}}(\frac{h^3_y}{h^7}-\frac{h_y}{h^5})kdk,\\
&\sigma^{\text{QM}}_{yxx}=-\frac{e^3}{4\pi^2\hbar}\sum_s\int f_s(2\frac{\partial G^{xx}_s}{\partial k_y}-\frac{\partial G^{xy}_s}{\partial k_x})d^2k\approx\frac{5\gamma^3e^3}{16\pi^2\hbar}\int^{2\pi}_0d\phi\int^{k_{F+}}_{k_{F-}}\frac{h^3_y}{h^7}kdk,\\
&\sigma^{\text{QM}}_{xxy}=-\frac{e^3}{4\pi^2\hbar}\sum_s\int f_s(\frac{3}{2}\frac{\partial G^{xy}_s}{\partial k_x}-\frac{1}{2}\frac{\partial G^{xx}_s}{\partial k_y})d^2k\approx\frac{5\gamma^3e^3}{16\pi^2\hbar}\int^{2\pi}_0d\phi\int^{k_{F+}}_{k_{F-}}(\frac{h^3_y}{h^7}-\frac{h_y}{h^5})kdk,\\
&\sigma^{\text{QM}}_{yxy}=-\frac{e^3}{4\pi^2\hbar}\sum_s\int f_s(\frac{3}{2}\frac{\partial G^{xy}_s}{\partial k_y}-\frac{1}{2}\frac{\partial G^{yy}_s}{\partial k_x})d^2k\approx\frac{5\gamma^3e^3}{16\pi^2\hbar}\int^{2\pi}_0d\phi\int^{k_{F+}}_{k_{F-}}(\frac{h^3_x}{h^7}-\frac{h_x}{h^5})kdk,\\
&\sigma^{\text{QM}}_{xyy}=-\frac{e^3}{4\pi^2\hbar}\sum_s\int f_s(2\frac{\partial G^{yy}_s}{\partial k_x}-\frac{\partial G^{yx}_s}{\partial k_y})d^2k\approx\frac{5\gamma^3e^3}{16\pi^2\hbar}\int^{2\pi}_0d\phi\int^{k_{F+}}_{k_{F-}}\frac{h^3_x}{h^7}kdk.\\
\end{aligned}
\end{equation}
To first order in $\Delta$, we have
\begin{equation}\label{aq-19}
\begin{aligned}
&\int^{2\pi}_0d\phi\int^{k_{F+}}_{k_{F-}}\frac{h^3_x}{h^7}kdk\approx\int^{2\pi}_0d\phi\int^{k_{F+}}_{k_{F-}}(\gamma^3k^3_x-3\gamma^2k^2_x\Delta\hat{m}_x)[\frac{1}{|\gamma|^7k^6}+\frac{7\gamma\Delta}{|\gamma|^9k^7}(\hat{m}_x\cos\phi+\hat{m}_y\sin\phi)]dk\\
&\approx\int^{2\pi}_0d\phi\int^{k_{F+}}_{k_{F-}}[\frac{\cos^3\phi}{\gamma^3|\gamma|k^3}+\frac{7\Delta\cos^3\phi}{|\gamma|^5k^4}(\hat{m}_x\cos\phi+\hat{m}_y\sin\phi)-\frac{3\Delta\hat{m}_x\cos^2\phi}{|\gamma|^5k^4}]dk\\
&=\int^{2\pi}_0[\frac{\cos^3\phi}{\gamma^3|\gamma|}(\frac{1}{2k^2_{F-}}-\frac{1}{2k^2_{F+}})+\frac{\Delta\hat{m}_x(7\cos^4\phi-3\cos^2\phi)+7\Delta\hat{m}_y\cos^3\phi\sin\phi}{|\gamma|^5}(\frac{1}{3k^3_{F-}}-\frac{1}{3k^3_{F+}})]d\phi\\
&\approx-\frac{2k_W(3k^2_0+4k^2_W)}{3k^6_0}\int^{2\pi}_0\frac{\Delta\hat{m}_x(7\cos^4\phi-3\cos^2\phi)}{|\gamma|^5}d\phi\\
&=-\frac{3\pi(3\hbar^2\epsilon_F+2m_e\gamma^2)\Delta}{8m_e\gamma^4\epsilon^3_F}\sin\theta\cos\varphi,
\end{aligned}
\end{equation}
and
\begin{equation}\label{aq-20}
\begin{aligned}
&\int^{2\pi}_0d\phi\int^{k_{F+}}_{k_{F-}}\frac{h_x}{h^5}kdk\approx\int^{2\pi}_0d\phi\int^{k_{F+}}_{k_{F-}}(\gamma k_x-\Delta\hat{m}_x)[\frac{1}{|\gamma|^5k^4}+\frac{5\gamma\Delta}{|\gamma|^7k^5}(\hat{m}_x\cos\phi+\hat{m}_y\sin\phi)]dk\\
&\approx\int^{2\pi}_0d\phi\int^{k_{F+}}_{k_{F-}}[\frac{\gamma\cos\phi}{|\gamma|^5k^3}+\frac{5\Delta\cos\phi}{|\gamma|^5k^4}(\hat{m}_x\cos\phi+\hat{m}_y\sin\phi)-\frac{\Delta\hat{m}_x}{|\gamma|^5k^4}]dk\\
&=\int^{2\pi}_0[\frac{\gamma\cos\phi}{|\gamma|^5}(\frac{1}{2k^2_{F-}}-\frac{1}{2k^2_{F+}})+\frac{\Delta}{|\gamma|^5}(5\hat{m}_x\cos^2\phi+5\hat{m}_y\cos\phi\sin\phi-\hat{m}_x)(\frac{1}{3k^3_{F-}}-\frac{1}{3k^3_{F+}})]d\phi\\
&\approx-\frac{2k_W(3k^2_0+4k^2_W)}{3k^6_0}\int^{2\pi}_0[\frac{\Delta\hat{m}_x}{|\gamma|^5}(5\cos^2\phi-1)]d\phi\\
&=-\frac{\pi(3\hbar^2\epsilon_F+2m_e\gamma^2)\Delta}{2m_e\gamma^4\epsilon^3_F}\sin\theta\cos\varphi,
\end{aligned}
\end{equation}
and
\begin{equation}\label{aq-21}
\begin{aligned}
&\int^{2\pi}_0d\phi\int^{k_{F+}}_{k_{F-}}\frac{h_y}{h^5}kdk\approx\int^{2\pi}_0d\phi\int^{k_{F+}}_{k_{F-}}(\gamma k_y-\Delta\hat{m}_y)[\frac{1}{|\gamma|^5k^4}+\frac{5\gamma\Delta}{|\gamma|^7k^5}(\hat{m}_x\cos\phi+\hat{m}_y\sin\phi)]dk\\
&\approx\int^{2\pi}_0d\phi\int^{k_{F+}}_{k_{F-}}[\frac{\gamma\sin\phi}{|\gamma|^5k^3}+\frac{5\Delta\sin\phi}{|\gamma|^5k^4}(\hat{m}_x\cos\phi+\hat{m}_y\sin\phi)-\frac{\Delta\hat{m}_y}{|\gamma|^5k^4}]dk\\
&=\int^{2\pi}_0[\frac{\gamma\sin\phi}{|\gamma|^5}(\frac{1}{2k^2_{F-}}-\frac{1}{2k^2_{F+}})+\frac{\Delta}{|\gamma|^5}(5\hat{m}_x\cos\phi\sin\phi+5\hat{m}_y\sin^2\phi-\hat{m}_y)(\frac{1}{3k^3_{F-}}-\frac{1}{3k^3_{F+}})]d\phi\\
&\approx-\frac{2k_W(3k^2_0+4k^2_W)}{3k^6_0}\int^{2\pi}_0[\frac{\Delta\hat{m}_y}{|\gamma|^5}(5\sin^2\phi-1)]d\phi\\
&=-\frac{\pi(3\hbar^2\epsilon_F+2m_e\gamma^2)\Delta}{2m_e\gamma^4\epsilon^3_F}\sin\theta\sin\varphi,
\end{aligned}
\end{equation}
and
\begin{equation}\label{aq-22}
\begin{aligned}
&\int^{2\pi}_0d\phi\int^{k_{F+}}_{k_{F-}}\frac{h^3_y}{h^7}kdk\approx\int^{2\pi}_0d\phi\int^{k_{F+}}_{k_{F-}}(\gamma^3k^3_y-3\gamma^2k^2_y\Delta\hat{m}_y)[\frac{1}{|\gamma|^7k^6}+\frac{7\gamma\Delta}{|\gamma|^9k^7}(\hat{m}_x\cos\phi+\hat{m}_y\sin\phi)]dk\\
&\approx\int^{2\pi}_0d\phi\int^{k_{F+}}_{k_{F-}}[\frac{\sin^3\phi}{\gamma^3|\gamma|k^3}+\frac{7\Delta\sin^3\phi}{|\gamma|^5k^4}(\hat{m}_x\cos\phi+\hat{m}_y\sin\phi)-\frac{3\Delta\hat{m}_y\sin^2\phi}{|\gamma|^5k^4}]dk\\
&\approx-\frac{2k_W(3k^2_0+4k^2_W)}{3k^6_0}\int^{2\pi}_0\frac{\Delta\hat{m}_y(7\sin^4\phi-3\sin^2\phi)}{|\gamma|^5}d\phi\\
&=-\frac{3\pi(3\hbar^2\epsilon_F+2m_e\gamma^2)\Delta}{8m_e\gamma^4\epsilon^3_F}\sin\theta\sin\varphi.
\end{aligned}
\end{equation}
with $k_0=\sqrt{2m_e\epsilon_F/\hbar^2}$ and $k_W=m_e|\gamma|/\hbar^2$. Substitution of Eqs. (\ref{aq-19})-(\ref{aq-22}) in Eq. (\ref{aq-18}) yields
\begin{equation}\label{aq-23}
\begin{aligned}
&\sigma^{\text{QM}}_{xxx}=\sigma^{\text{QM}}_W\sin\theta\cos\varphi, \sigma^{\text{QM}}_{yyy}=\sigma^{\text{QM}}_W\sin\theta\sin\varphi, \sigma^{\text{QM}}_{yxx}=3\sigma^{\text{QM}}_W\sin\theta\sin\varphi,\\
&\sigma^{\text{QM}}_{xxy}=-\sigma^{\text{QM}}_W\sin\theta\sin\varphi, \sigma^{\text{QM}}_{yxy}=-\sigma^{\text{QM}}_W\sin\theta\cos\varphi, \sigma^{\text{QM}}_{xyy}=3\sigma^{\text{QM}}_W\sin\theta\cos\varphi,\\
&\sigma^{\text{QM}}_W\equiv-\frac{5e^3(3\hbar^2\epsilon_F+2m_e\gamma^2)\Delta}{128\pi\hbar m_e\gamma\epsilon^3_F}.
\end{aligned}
\end{equation}
We substitute in Eq. (\ref{eq-10}) and find that
\begin{equation}\label{aq-24}
\sigma^{\text{QM}}_{\phi\phi\phi}=\sigma^{\text{QM}}_{xxx}\cos\phi+\sigma^{\text{QM}}_{yyy}\sin\phi=\sigma^{\text{QM}}_W\sin\theta\cos(\varphi-\phi).
\end{equation}
For PSO, we have $\mathbf{h(k)}=(\delta(k_x-k_y)-\Delta\hat{m}_x, \delta(k_x-k_y)-\Delta\hat{m}_y, -\Delta\hat{m}_z)$ and
\begin{equation}\label{aq-25}
G^{xx}_s=G^{yy}_s=-G^{xy}_s=s\delta^2\frac{2h^2-(h_x+h_y)^2}{4h^5}.
\end{equation}
We substitute in Eq. (\ref{eq-11}) and find that
\begin{equation}\label{aq-26}
\begin{aligned}
&\sigma^{\text{QM}}_{xxx}=-\frac{e^3}{4\pi^2\hbar}\sum_s\int f_s\frac{\partial G^{xx}_s}{\partial k_x}d^2k=-\frac{5\delta^3e^3}{16\pi^2\hbar}\int^{2\pi}_0d\phi\int^{k_{F+}}_{k_{F-}}[\frac{(h_x+h_y)^3}{h^7}-\frac{2(h_x+h_y)}{h^5}]kdk,\\
&\sigma^{\text{QM}}_{yyy}=-\frac{e^3}{4\pi^2\hbar}\sum_s\int f_s\frac{\partial G^{yy}_s}{\partial k_y}d^2k=-\sigma^{\text{QM}}_{xxx},\\
&\sigma^{\text{QM}}_{yxx}=-\frac{e^3}{4\pi^2\hbar}\sum_s\int f_s(2\frac{\partial G^{xx}_s}{\partial k_y}-\frac{\partial G^{xy}_s}{\partial k_x})d^2k=-\sigma^{\text{QM}}_{xxx},\\
&\sigma^{\text{QM}}_{xxy}=-\frac{e^3}{4\pi^2\hbar}\sum_s\int f_s(\frac{3}{2}\frac{\partial G^{xy}_s}{\partial k_x}-\frac{1}{2}\frac{\partial G^{xx}_s}{\partial k_y})d^2k=-\sigma^{\text{QM}}_{xxx},\\
&\sigma^{\text{QM}}_{yxy}=-\frac{e^3}{4\pi^2\hbar}\sum_s\int f_s(\frac{3}{2}\frac{\partial G^{xy}_s}{\partial k_y}-\frac{1}{2}\frac{\partial G^{yy}_s}{\partial k_x})d^2k=\sigma^{\text{QM}}_{xxx},\\
&\sigma^{\text{QM}}_{xyy}=-\frac{e^3}{4\pi^2\hbar}\sum_s\int f_s(2\frac{\partial G^{yy}_s}{\partial k_x}-\frac{\partial G^{yx}_s}{\partial k_y})d^2k=\sigma^{\text{QM}}_{xxx}.\\
\end{aligned}
\end{equation}
It is seen that only $\sigma^{\text{QM}}_{xxx}$ needs to be calculated. Moreover, from the Fig. \ref{f-2}(f), the integrand $\partial G^{xx}_s/\partial k_x$ is pronounced around $k_x=k_y$. Thus, we have
\begin{equation}\label{aq-27}
\begin{aligned}
&\sigma^{\text{QM}}_{xxx}=-\frac{5\delta^3e^3}{16\pi^2\hbar}\int^{2\pi}_0d\phi\int^{k_{F+}}_{k_{F-}}[\frac{(h_x+h_y)^3}{h^7}-\frac{2(h_x+h_y)}{h^5}]kdk,\\
&\approx-\frac{5\delta^3e^3}{16\pi^2\hbar}\int^{2\pi}_0d\phi\int^{k_{F+}}_{k_{F-}}[\frac{8\delta^3k^4(\cos\phi-\sin\phi)^3}{|\Delta|^7}-\frac{12\Delta\delta^2k^3(\cos\phi-\sin\phi)^2\sin\theta(\cos\varphi+\sin\varphi)}{|\Delta|^7}\\
&-\frac{4\delta k^2(\cos\phi-\sin\phi)}{|\Delta|^5}+\frac{2\Delta k\sin\theta(\cos\varphi+\sin\varphi)}{|\Delta|^5}]dk\\
&\approx-\frac{5\delta^3e^3\sin\theta(\cos\varphi+\sin\varphi)}{16\pi^2\hbar}\int^{2\pi}_0d\phi\int^{k_{F+}}_{k_{F-}}[\frac{2\Delta k}{|\Delta|^5}-\frac{12\Delta\delta^2k^3(\cos\phi-\sin\phi)^2}{|\Delta|^7}]dk\\
&\approx-\frac{5\delta^3e^3\Delta\sin\theta(\cos\varphi+\sin\varphi)}{8\pi^2\hbar|\Delta|^5}(-\frac{16\sqrt{2m^3_e\epsilon_F}|\delta|}{\hbar^3}-\frac{32\sqrt{2m^5_e}|\delta|^3}{3\hbar^5\sqrt{\epsilon_F}}+\frac{256\sqrt{2m^5_e\epsilon_F^3}|\delta|^3}{\hbar^5\Delta^2}
+\frac{3072\sqrt{2m^7_e\epsilon_F}|\delta|^5}{3\hbar^7\Delta^2}+\frac{24576\sqrt{2m^9_e}|\delta|^7}{35\hbar^9\Delta^2\sqrt{\epsilon_F}})\\
&=-\frac{\sigma^{\text{QM}}_P}{2\sqrt{2}}\sin\theta\sin(\varphi+\frac{\pi}{4}).
\end{aligned}
\end{equation}
In Eq. (\ref{aq-27}), we have used the zero-order approximation for $k_{Fs}$, that is, $k_{Fs}\approx\sqrt{k_\phi^2+k_0^2}-sk_\phi$ with $k_\phi=\sqrt{2}m_e|\delta(\text{cos}\phi-\text{sin}\phi)|/\hbar^2$ and $k_0=\sqrt{2m_e\epsilon_F/\hbar^2}$. We substitute in Eq. (\ref{eq-10}) and find that
\begin{equation}\label{aq-28}
\begin{aligned}
\sigma^{\text{QM}}_{\phi\phi\phi}&=\sigma^{\text{QM}}_{xxx}\cos^3\phi-\sigma^{\text{QM}}_{xxx}\sin^3\phi+3\sigma^{\text{QM}}_{xxx}\sin^2\phi\cos\phi-3\sigma^{\text{QM}}_{xxx}\sin\phi\cos^2\phi\\
&=\sigma^{\text{QM}}_{xxx}(\cos\phi-\sin\phi)^3=-2\sqrt{2}\sigma^{\text{QM}}_{xxx}\sin^3(\phi-\frac{\pi}{4})\\
&\approx\sigma^{\text{QM}}_P\sin\theta\sin(\varphi+\frac{\pi}{4})\sin^3(\phi-\frac{\pi}{4}).
\end{aligned}
\end{equation}
\end{widetext}

\end{document}